\documentstyle[prl,aps]{revtex} 
\draft

\begin{document}
\twocolumn[\hsize\textwidth\columnwidth\hsize\csname
@twocolumnfalse\endcsname
\title{%
\hbox to\hsize{\normalsize\rm April 1997\hfil 
Preprint MPI-PTh/97-28}
\bigskip\bigskip
Comment on ``Cherenkov Radiation by Neutrinos in a
Supernova Core''}
\author{Georg G.~Raffelt}
\address{Max-Planck-Institut f\"ur Physik,
F\"ohringer Ring 6, 80805 M\"unchen, Germany}
\date{16 April 1997}
\maketitle
\begin{abstract}
Mohanty and Samal have shown that the magnetic-moment interaction with
nucleons contributes significantly to the photon dispersion relation
in a supernova core, and with an opposite sign relative to the usual
plasma effect. Because of a numerical error they overestimated the
magnetic-moment term by two orders of magnitude, but it is still of
the same order as the plasma effect. It appears that the Cherenkov
processes $\gamma\nu\to\nu$ and $\nu\to\nu\gamma$ remain forbidden,
but a final verdict depends on a more detailed investigation of the
dynamical magnetic susceptibility of a hot nuclear medium.
\end{abstract}
\vskip2.0pc]

Mohanty and Samal \cite{MS96} found that in a supernova (SN) core the
photon refractive index receives a contribution from the nucleon
magnetic moments which is far more important and of opposite sign
compared with the usual electronic plasma effect. This would cause the
photon four momentum $K=(\omega,{\bf k})$ to be spacelike, allowing
for the Cherenkov processes $\nu\to\nu\gamma$ or $\gamma\nu\to\nu$
with important consequences for the neutrino opacities or the
emissivities of right-handed neutrinos. Unfortunately, a huge
numerical error has crept into their analysis, invalidating their
conclusions. 

The contribution of the nucleon magnetic moments $\mu$ to the photon
refractive index is $n_{\rm refr}^2=1+\chi$ in terms of the magnetic
susceptibility $\chi$. (Mohanty and Samal used unrationalized units
where $4\pi\chi$ appears instead.) For small photon frequencies
$\omega\to0$ and in the long-wavelength limit ${\bf k}\to0$ one may
use the static Pauli susceptibility of the nucleons which in the
degenerate limit is $\chi_0=\mu^2 m_N \,p_F/\pi^2$ with $m_N$ the
nucleon mass and $p_F$ its Fermi momentum.  Expressing the magnetic
moment in the usual form $\mu=\kappa\,e/2m_N$ gives us $n_{\rm
refr}^2-1=\kappa^2\,(\alpha/\pi)\,(p_F/m_N)$ with
$\alpha=e^2/4\pi\approx 1/137$ the fine-structure constant. For the
conditions assumed in Ref.~\cite{MS96} the contributions from protons
($\kappa_p=2.79$) and neutrons ($\kappa_n=-1.91$) add to $n_{\rm
refr}^2-1=0.81\times10^{-2}$, a factor $10^{-2}$ smaller than in
Ref.~\cite{MS96}. Mohanty and Samal agree about this correction
(private communication). 

Intruigingly, this reduced result is still of the same order as the
electronic contribution which is well approximated by $n_{\rm
refr}^2-1=-\frac{3}{2}\,\omega_P^2/\omega^2$. In a SN core the
electrons are degenerate so that the plasma frequency is
$\omega_P^2=(4\alpha/3\pi) p_F^2$ with $p_F$ the electron Fermi
momentum~\cite{Braaten}, leading to $n_{\rm
refr}^2-1=-(2\alpha/\pi)\,(p_F/\omega)^2$.  Because in a SN core the
nucleons are only partially degenerate it is justified to compare this
plasma term with the magnetic-moment contribution of nondegenerate
nucleons. I find $\chi_0=\mu^2 n_N/T$, leading to $n_{\rm
refr}^2-1=\kappa^2\,(\alpha/3\pi) \,(p_F^3/Tm_N^2)$ where
$n_N=p_F^3/3\pi^2$ was used for the nucleon density.  Summing the
contributions of protons and neutrons I find
\begin{eqnarray}\label{eq:comparison}
\frac{(n_{\rm refr}^2-1)_{P}}
{(n_{\rm refr}^2-1)_{\mu}}
&=&-f(Y_e)\,\frac{T m_N^2}{2\omega^2 p_F}\\
&&\hskip-6.2em
=\,-6.0\,f(Y_e)\,\frac{T}{30\,\rm MeV}\,
\left(\frac{100\,\rm MeV}{\omega}\right)^2
\left(\frac{10^{14}\,\rm g/cm^3}{\rho}\right)^{1/3}.
\nonumber
\end{eqnarray}
Here, $\rho=m_Nn_N$ is the mass density while the ``baryon Fermi
momentum'' $p_F=(3\pi^2 n_N)^{1/3}$ is just a parameter to
characterize the density. Finally,
$f(Y_e)\equiv12\,Y_e^{2/3}/[Y_e\kappa_p^2+(1-Y_e)\kappa_n^2]$
is a slowly varying function which is about unity for a typical
electron/baryon number fraction $Y_e=0.3$. The proton fraction has
been taken to equal $Y_e$ because of charge neutrality so that there
are $1-Y_e$ neutrons per baryon.

Typical photons have energies around $\omega=3T$ so that even
$T=60~\rm MeV$ and $\rho=10^{15}~\rm g/cm^3$ will not bring the ratio
below unity.  It appears that for all conditions occurring in a SN
core the plasma term dominates, but only by a surprisingly narrow
margin.

Because the plasma term decreases with $\omega^{-2}$ one may think
that for the high-energy tail of the photon distribution the
magnetic-moment term could dominate. One needs to remember, though,
that this term is based on a thermodynamic derivation in the static
limit and thus is valid only in the hydrodynamic limit, i.e.\ for
frequencies well below the spin relaxation rate.

The spin relaxation rate in a SN core is not known, but probably does
not exceed the temperature by much~\cite{Janka}. For frequencies
exceeding the spin relaxation rate the spin susceptibility decreases,
and should in fact become negative for large frequencies as can be
shown by virtue of the Kramers-Kronig relations and the f-sum rule for
the spin-density structure function. For large $\omega$ I find
$n_{\rm refr}^2-1 \propto -\omega^{-2}$, i.e.\ the same frequency
dependence of the refractive index as that of the plasma effect. Note
that for noninteracting fermions the magnetic-moment refractive index
vanishes because the photon forward-scattering amplitude on a fermion
with magnetic moment vanishes. Therefore, for frequencies far
exceeding the spin relaxation rate the magnetic-moment induced
refractive index must indeed vanish.  These matters will be discussed
in more detail elsewhere.

As it stands, I believe that even though the plasma and nucleon
magnetic moment contributions to the photon refractive index are
tantalizingly close to each other, the plasma effect seems to win for
all plausible conditions of a SN core.  Therefore, the neutrino
Cherenkov effect does not seem to occur in a SN core.  However,
because the plasma and magnetic-moment effects are so close to each
other a more detailed investigation is called for.

I thank S.~Mohanty for a collegial correspondence and P.~Elmfors,
A.~Kopf, and G.~Sigl for helpful remarks on an early version of the
manuscript. 

\newpage

This work was supported, in part, by the Deutsche
Forschungsgemeinschaft under grant No.\ SFB 375.


\end{document}